# Personal Investigator: a Therapeutic 3D Game for Teenagers

**David Coyle**
HEA Researcher/Computer Science Dept.
Media Lab Europe/Trinity College Dublin
davidc@medialabeurope.org
+353 86 8269863

**Mark Matthews**
HEA Researcher/Computer Science Dept.
Media Lab Europe/Trinity College Dublin
mark@medialabeurope.org
+353 86 1524921

**ABSTRACT**
This position paper describes the implementation and initial findings of a game called Personal Investigator (PI). PI is an online 3D detective game that implements a model of Brief Solution Focused Therapy (BSFT). It aims to help teenagers overcome mental health problems and engage with traditional mental health care services. It is predicted that the combination of goal-oriented gaming with a model of goal-oriented therapy will help to attract and sustain the interest of teenagers, a group that therapists often have difficulty engaging with. PI is the first game to integrate this established psychotherapy approach into an engaging online 3D game.

**INTRODUCTION**
PI is an initial year one demo of a three-year project to test the suitability of online 3D games for providing positive mental health care to teenagers. In PI teenagers play the role of a Solution Detective and move around the Detective Academy. As they journey through the academy they meet several characters and are set a series of tasks. Rewards are given as each task is achieved and if the teenager succeeds at all the tasks they are able to graduate from the academy and become a Master Detective. The tasks and dialogues in the game implement a model of BSFT. This form of therapy is particularly suitable for implementation within a gaming model because it is goal oriented. By achieving the goals set by the game, teenagers learn to clarify their problem, convert their problem to a goal and identify the resources they have to help them achieve this goal.

The game is currently undergoing initial trials in the Department of Child and Family Psychiatry at the Mater Hospital Dublin. The duration of the trials will be December 2003 to March 2004. Professional psychotherapists will test the game with groups of teenagers and assess its impact and effectiveness. We aim to answer the following questions:

- Can a model of BSFT be successfully integrated into a 3D computer game?
- Can such a game improve a therapist's ability to engage with teenagers?
- Does the game help teenagers to overcome emotional problems and if so how specifically does it help?
- What level of gameplay is appropriate in such a game?

**INCORPORATING SUBJECT MATTER EXPERTS**
The great strength of this game is that it is a coming together of traditional psychiatric services and computer game researchers. Dr John Sharry, Principal Social Worker in the Department of Child and Family Psychiatry Mater Hospital, is a consultant researcher and developer on the project and is organising the clinical trials of PI. Dr Sharry has written several books on models of BSFT. The game is one of several being developed by the Mindgames group at Media Lab Europe, the European research partner of MIT Media Lab. The group is developing an approach called Affective Feedback, a combination of intelligent biofeedback, video gaming and sensory immersion to positively affect a participant's state of mind.

**THE CONTEXT AND OPPORTUNITY**
Although mental health problems increase markedly during teenage years, therapists often find it difficult to engage with teenagers. The majority of disturbed teenagers do not receive professional mental health care and of those who do fewer still will fully engage with the therapeutic process [6, 9]. Disturbed teenagers are more likely to seek help from informal sources such as friends.

Teenagers are generally more private and self-conscious and also more confrontational than either younger children or adults. PI was developed because specific opportunities were identified:

- Therapists engage best with teenagers if their approach is teenager centred [8].
- Playing games can improve the therapeutic process and break down formal boundaries [8].

- Teenagers are comfortable with computer technologies and use the Internet to seek help for emotional problems.

Up to twenty percent of teenagers use the Internet to seek help for emotional problems [4]. Examples of the types of problems include problems with friends, family problems, academic problems and suicidal thoughts. Seeking help from the Internet overcomes the gender divide. Boys, who are usually significantly less likely to seek help, are just as likely to seek Internet help as girls [4]. This is particularly important because it suggests the Internet can be used to target boys, who are resistant to existing prevention methods for high-risk behaviours such as suicide.

One major concern that exists about current Internet help seeking is that, despite the growing number of websites underwritten by health care organisations, the most popular source of information is chat rooms [4]. Chat rooms are the Internet equivalent of informal help sources, but with the added risk that instead of approaching a trusted friend or family member, you are approaching a stranger whose expertise and intentions (positive or malicious) are unknown [1]. Mental health care organisations need to address this issue. Research is necessary to make empirical sources of online help more attractive to teenagers. A hugely important fact about Internet help seeking is that disturbed teenagers who find beneficial help online are subsequently more likely to seek formal help from traditional mental health care services [4].

When people do approach traditional mental health care services the outcome is most likely to be successful if the therapist engages with the client in a client centred way. For adults dialogue is the favoured means of communication. This is not the case with children. Children relate better through gesture and play. Much research has been conducted into ways of engaging children in a therapeutic process. Some examples of tools used are storybooks, construction materials, artwork and puppets. Therapists can engage with children by playing children's games. Teenagers are resistant to these methods, they like to be treated as adults and will not engage with a therapist if they perceive they are being treated as a child. Equally however many teenagers are private and self-conscious and often react confrontationally or not at all to direct dialogue with a therapist. Worksheets or questionnaires are the tools frequently used to overcome this barrier [8]. The aim of our research is to develop more powerful tools (for teenagers and therapists), drawing on the lessons learned about game play with younger children.

Our group is developing a computer-aided model of how therapists can engage with teenagers (fig.1). Instead of engaging directly with a teenager the therapist uses a computer as a third party in their dialogue. This makes the dialogue less confrontational and increases the likelihood of teenagers talking about awkward personal issues.

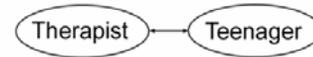
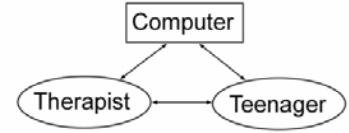

**Figure 1. Computer aided communication.**

The computer game is a powerful medium because it is an accepted part of teenage culture.

Given the context outlined above, we had three main aims in developing PI:

- To design a 3D computer game that helps teenagers to solve personal problems.

- To design a 3D computer game that enables therapists to engage more easily with troubled teenagers.

- To design a 3D computer game that is accessible online but can also be used in one to one sessions.

## BRIEF SOLUTION FOCUSED THERAPY AND GAMING

BSFT is an established and effective strengths-based, goal-focused approach to counseling and psychotherapy. Evidence shows that BSFT is as effective as traditional psychotherapies in helping clients [3, 5]. BSFT helps clients construct solutions rather than to focus on problems, concentrating on the future and not on the past [2, 7]. The approach can be divided into five therapeutic conversational strategies:

*Setting Goals*
In BSFT clients do not talk about problems, they set goals they want to achieve, elaborating on them in as much detail as possible.

*Recognising Exceptions*
Exceptions are times when the client's problem is not present or is less acute. BSFT helps clients recognise and explore these times with a view to repeating them more often.

*Coping*
BSFT helps clients to recognise ways they currently have of dealing with their problem and how they have successfully overcome past problems.

*Identifying resources*
BSFT helps clients identify resources, in particular support from family and friends, which they can draw upon. Drawing on this support can make a vital difference for clients. Resources refers also to the clients own strengths i.e. things they are good at.

*The Miracle Question*
"Imagine you woke up tomorrow and the problem was solved, how would your life be different?" By imagining a future without their problems, clients are motivated to seek a solution.

BSFT was chosen as the therapeutic basis for PI because it shares a goal-oriented approach with computer games. Both actively use goals as a form of motivation. In a BSFT session the therapist and client set a goal they want to achieve. Computer games operate in a similar way. Players must achieve minor goals (e.g. fight a beast, vault a wall) to achieve the major goals (e.g. finish the game). In both BSFT and computer games new skills or strategies must be learned in order to achieve goals. Furthermore, the skills learnt achieving a goal can often be reused or adapted to attain further goals. A traditional computer game will reward its players for reaching a goal in different ways, e.g. receiving a special new tool, reaching the next level. In PI the goals defined for the game are therapeutic goals, which will benefit the client in their day-to-day life.

**THE GAME - PERSONAL INVESTIGATOR**

In PI teenagers play the role of a Solution Detective and visit the Detective Academy. The overall goal is to graduate from the Detective Academy as a Master Detective. To implement BSFT into a game, it was broken down into the five therapeutic strategies described above. These strategies were then mapped into five game areas (figure 2.).

Players begin by logging in with a username and password. An individual account is created which allows players to save their progress and return at a later point. Players are explicitly asked whether they would like to play and must answer 'yes' to continue. This is a small but important therapeutic step, as it acknowledges a positive desire to solve a problem.

Once the game begins, the player is transported into the game world and starts in the Introduction Area in front of the Detective Academy. The first character the player meets is the principal of the Detective Academy. He invites the teenager to set a therapeutic goal i.e. to choose a problem they want to solve. This character also gives the player a detective notebook, which appears at the bottom of the screen, and is used throughout the game to record goals, resources and thoughts.

Inside the Detective Academy, there are four distinct areas to be explored, which correspond to the four remaining aspects of BSFT (figure 2.). Waiting in each of these areas is a master detective who assigns the player tasks and rewards their completion with a key. For example in the 'Backup' area, the player meets an American policeman, who specialises in identifying resources. He helps players understand the need for support and invites them to watch a video testimony from one of his former pupils. This pupil tells the player how speaking to her brother helped her get over her parents separation. When players identify their own strengths and resources, they receive a key and can move on to the next area. Players must collect a key from each area. All four keys are required in order to graduate from the academy. As a reward for graduation, players receive a printout of the detective notebook, which can then be used in further therapeutic work outside of the game.

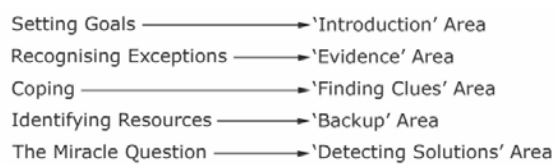

**Figure 2. Mapping the game areas.**

PI delivers high quality graphics to an online environment. It was constructed using two main authoring packages. A beta version of Atmosphere, Adobe's new virtual environment package, was used to create the 3D environments. Atmosphere is quick to develop in, its lighting is excellent and it runs online through Internet Explorer. The game was programmed using the Atmosphere JavaScript API. The detective notebook was constructed using Macromedia Flash MX. Atmosphere and Flash communicate via the web browser.

**DISCUSSION OF INITIAL FINDINGS**

At the time of writing PI is undergoing initial trials in the Mater Hospital in Dublin. These trails will involve trained therapists using the game with referred teenagers, gaining questionnaire feedback from both parties on completion. Initial results will be available in April 2004 in time for the CHI2004 conference. Below we discuss some of the tentative conclusions we can make based on initial questionnaire feedback from therapists and games developers who have completed the game.

*Therapeutic Aspect*

Initial testing indicates that the use of 3D is appropriate. The use of a virtual world is powerful because it offers troubled teenagers somewhere safe to go to address their problems. In the virtual world they can more easily forget the troubles that beset their daily lives and focus on identifying new and existing strengths.

Attending therapeutic sessions is difficult for teenagers. Indications are that the 3D game will add to the absorption of the teenager in a therapeutic session. The use of music and sound effects makes the task more entertaining. A single discussion with a therapist is replaced by five shorter task-based discussions with five distinctive characters. This helps to prevent boredom. Visual rewards and encouragement given by the game reinforce the sense of progress being made.

The 3D environment allows for pacing within the therapeutic session. The teenager plays the game at their own pace, over one or many sessions. At any point the teenager can stop and talk to the therapist to further discuss ideas generated by the game. The therapist can observe unobtrusively and choose to have as much or as little involvement in the game as seems beneficial.

Being able to print a record of your detective notebook is a positive aspect of the game. The printout acts both as a reward for completing the game and as a permanent record of the lessons learnt in the game.

PI has implemented a model of BSFT in an open manner. The game is not targeted at a specific problem e.g. substance abuse, bullying, family traumas. The game asks the teenager what problem they would like to solve and is designed to be flexible enough to help solve whatever problem they define. Therapists surveyed have indicated that they would like to be able to easily modify the dialogue in the game to address specific issues on a case-by-case basis.

An open question in the development of PI was what degree of gameplay the game should have? Game design researchers wanted to increase the amount of gameplay whereas therapists preferred to reduce gameplay and focus on the dialogues implemented by the game. Therapists feared that too much gameplay might distract teenagers from the therapeutic goals. Game designers argued that as the gameplay would revolve around achieving therapeutic goals, it would increase teenagers involvement in and enjoyment of the therapeutic process. It was decided to err on the side of caution and reduce gameplay. A comparative study is necessary, with a future implementation increasing the level of gameplay.

*Technical Aspect*

PI has been successfully implemented and tested as an online game. The download size of the game (approximately 19MB) dictates that it is only possible to play the game using a broadband Internet connection. The use of Adobe Atmosphere has meant that the game is graphically rich. A future implementation of the game with higher degrees of gameplay will demand the use of a more specialised games engine.

**CONCLUSION**

As we have not completed trials of PI, it is not yet possible to make any firm conclusions from our research. However, based on initial surveys, we can say that PI demonstrates that BSFT can be successfully implemented into a 3D computer game. We can tentatively conclude that PI helps engage teenagers in therapeutic sessions. It could also help a therapist build up a relationship with an adolescent client. Further tests are necessary to determine whether PI can engage teenagers online, whether its amount of gameplay is suitable and whether it can help teenagers overcome mental health problems both online and in therapy.

**DEMONSTRATING**

PI is available for demonstration and can be made available in advance of the workshop for testing.

**BIOGRAPHIES**

Mark Matthews and David Coyle have worked together on several projects. Both are joint HEA Researchers with Media Lab Europe and PhD candidates with the Computer Science Department, Trinity College Dublin. Their current research concerns the use of computer games to develop therapeutic relationships, as well as investigating the learning possibilities of collaborative gaming. Both hold a first class MSc in Multimedia from Dublin City University. Their previous project Savant, a mixed media exploration of autistic experience, won the 2003 Europrix Award for the best DVD/CDROM produced in Europe in 2003. Their current project PI is short listed for an O2 Digital Media Award for Digital Innovation. Mark Matthews holds a BA in English and Philosophy from Trinity College Dublin. David Coyle holds a BSc in Electronic Engineering.

**ACKNOWLEDGMENTS**

We would like to thank Dr Andy Nisbet, our PhD supervisor in the Computer Science Department of Trinity College Dublin, for his help and work on the development of PI. We would also like to thank Gary McDarby, the head of the Mindgames group in Media Lab Europe (www.medialabeurope.org), and all the members of the Mindgames group. Thank you also to Dr. Louise Atkin.